\documentstyle[aps,prl,multicol,psfig]{revtex}
\tighten
\begin{document}

%
%%%%%%%%%%%%%%%%%%%%%%%%%%%%%%%%% TITLE PAGE
%
\title{Joint superexchange--Jahn-Teller mechanism for A-type
antiferromagnetism in ${\rm LaMnO_3}$}
\author{D. Feinberg$^\dagger$, P. Germain$^*$, M. Grilli$^{*,\dagger}$,
and G. Seibold$^*$}
\address{$^*$Istituto di Fisica della Materia e
Dipartimento di Fisica, Universit\`a di Roma ``La Sapienza'',\\
Piazzale A. Moro 2, 00185 Roma, Italy}
\address{$^\dagger$Laboratoire 
des Propri\'et\'es Electroniques des Solides, CNRS \\
25 Avenue des Martyrs, BP 166, 38042 Grenoble Cedex 9, France}
\maketitle
%
%%%%%%%%%%%%%%%%%%%%%%%%%%%%%%%%% ABSTRACT
%
\begin{abstract} 
We propose a mechanism for A-type antiferromagnetism in 
orthorombic ${\rm LaMnO_3}$, compatible with the large 
Jahn-Teller splitting inferred from structural data.
Orbital ordering resulting from Jahn-Teller distortions
effectively leads to A-type ordering (antiferromagnetic in the
$c$ axis and ferromagnetic in the $ab$ plane) provided the
in-plane distorsion $Q_2$ is large enough, a condition generally
fulfilled in existing data. 

\end{abstract}
%
%%%%%%%%%%%%%%%%%%%%%%%%%%%%%%%%% PACS NUMBERS
%
%
%%%%%%%%%%%%%%%%%%%%%%%%%%%%%%%%% PAPER BODY
%


\begin{multicols}{2}
Stoichiometric  ${\rm La Mn O_3}$ (LMO) 
is known \cite{wollan} to be an $A$-type antiferromagnetic insulator
(A-AFMI), where ferromagnetically
ordered ${\rm Mn O_2}$ planes (in the $xy$ direction)
have staggered magnetization along the $z$ axis. Upon increasing
the temperature a paramagnetic insulating phase (PMI) is reached.
On the other hand
sufficient hole doping (e.g. by substituting La with Sr or Ca)
gives rise via the so-called double-exchange hopping mechanism
\cite{zener,anderson}
to a low-temperature ferromagnetic metallic phase (FMM)
turning into a PMI phase at higher temperature.
Not only magnetism determines the main physical properties, in fact
both theoretical \cite{millis,bishop} and experimental \cite{expphonon}
evidences emphasize the relevance of electron-lattice coupling.
Charge and orbital ordering also occur, further showing the 
competition between various  physical mechanisms.
Notice that the crucial role
of spin and lattice coupling was repeatedly emphasized to account for
the properties of the FMM phase and the FMM-PMI transition at finite
doping, as well as charge-ordering phenomena.
However, this {\it liaison} regarded the double-exchange 
mechanism for charge tranport, being dynamically dressed by lattice
degrees of freedom \cite{millis,bishop,expphonon}. No emphasis
was put on the role of static 
cooperative Jahn-Teller (JT) deformations in stabilizing
specific magnetic structures in the AFMI phase.

Here we investigate an approach to the
stoichiometric phase of LMO showing that the layered antiferromagnetic
structure may result from the interplay between superexchange
and JT couplings. Our analysis is alternative
to the more qualitative one based on the semicovalent exchange mechanism
\cite{goodenough} and is complementary to the superexchange 
mechanism investigated by Kugel and Khomskii (KK) \cite{KK}
for perovskites with JT ions. This latter analysis (see also \cite{lacroix})
focused on the interplay between magnetic and orbital ordering
within the two $e_g$ orbital manifold, assumed degenerate. In particular, basic
ingredients were a strong local electron-electron repulsion $U$,
the Hund coupling $J_H$ between electrons on the two 
$e_g$ orbitals, and the orbital mixing (described by a mixing 
angle $\theta$) due to JT distorsion.
In the approach of Ref.\cite{KK}, only $e_g$ (spin and orbital)
degrees of freedom were considered, and
the spin and orbital order were self-consistently determined to lowest order in 
$J_H/U$.
The $e_g$ level degeneracy was lifted
by superexchange but the JT splitting resulting from the lattice distorsions
induced by orbital ordering was not explicitly considered, Only a 
correction due to a small local JT anharmonicity was introduced. This
point of view, considering magnetic exchange to be the main cause of
orbital mixing/ordering, 
but neglecting the orbital splitting resulting from JT effect,
might be questioned in LMO where strong JT 
distorsions arise. Moreover, magnetic exchange 
interactions are somewhat modest,
e.g. $J_{AF} \approx 0.58$ meV 
and $J_F \approx 0.83$ meV $ > J_{AF}$ from inelastic neutron 
scattering experiments \cite{moussa},
while the KK theory results in $J_F \approx (J_H/U)J_{AF} $, that is $J_F$ 
much smaller than $J_{AF}$. Moreover, as pointed out by KK, the observed 
orthorhombic distorsion with $c<a$ is not expected from 
considering solely orbital ordering. Therefore, 
it is not obvious, considering the actual distorsion, that the magnetic 
$A$-phase is still the most stable.

According to the experimental evidences for the relevance of 
the $t_{2g}$ (spin) degrees of freedom, e.g. in the double exchange
hopping processes, and for a strong JT coupling, 
we propose to reconsider the problem. We take properly into account 
the Hund coupling between $e_g$ 
and $t_{2g}$ electrons, and assume, contrarily to KK,
that the JT splitting is much larger than the 
exchange energy. This makes the JT effect the driving mechanism
for orbital ordering, which in turn controls the magnetic interactions.
Hereafter we {\it assume} some staggered 
orbital order (i.e. we fix $\theta$ on the two
sublattices) as determined by  strong
JT distorsions, without attempting to calculate them, since they can be
extracted from crystallographic data. Given the orbital order and the related
(relatively large) $e_g$ orbital splitting
$\epsilon$, we calculate at any order in $\epsilon$ and $J_H/U$
the superexchange interactions. Finally, we determine  the parameter ranges
which are compatible with the observed A-type AF phase. 

For the sake of simplicity we disregard the oxygen sites in the
perovskite structure, thus focusing on a single-site model.
On each site two Manganese orbitals, the $d_{x^2-y^2}$ ($x$)
and the $d_{3z^2-r^2}$ ($z$), are available. The real lattice
structure is effectively taken into account via the sign and 
the magnitude of the intersite hopping along the $x-y-z$ directions.
Specifically, we notice that, for a standard choice of phases, 
the hopping between the $d_x$ and the $d_z$
orbitals on Mn are given by
\begin{eqnarray}
t_{xx}=3t; & \;\;\; t_{xz}& =-\sqrt{3}t\,\, 
{\rm along} \;\;\hat{\bf x} \nonumber \\
t_{zz}=-t; & \;\;\; t_{xz}& =\sqrt{3}t,\,\, 
{\rm along} \;\;\hat{\bf y} \nonumber \\
t'_{zz}=-4t&\;\;\; t'_{xx}& =t'_{xz}=0 \;\;\; .
\label{xzhopping}
\end{eqnarray}
Throughout this paper
the apex indicates hoppings in the $z$ direction. The
(static) JT distortions \cite{yamada} mix the $x$ and $z$ orbitals
\cite{bishop} into $a$ (lower)
and $b$ (upper) orbitals with $e_g$ symmetry split by an
energy $\epsilon\equiv 2g\sqrt{Q_1^2+Q_2^2}$ where $g$ is an electron-lattice 
coupling constant. Specifically, using the same notations of Ref. 
\onlinecite{bishop}
the uniform distortion $Q_1$ (corresponding to a uniform variation of the
lattice parameter along the $z$ direction) couples to the 
$x-z$ density difference $n_x-n_z$. 
In systems like LMO, where the lattice spacing in the
$z$ direction is shorter that the (average) spacing in the $xy$ plane,
the $d_z$ orbitals overlap more and are pushed at higher energy
by the Coulomb interaction. Then $Q_1$ is negative. As pointed out by KK, this 
disfavours hopping in the z-direction, thus making the JT effect {\it compete}
with the A-type superexchange which alone would imply the opposite distortion.
The reverse
is true in systems like ${\rm KCuF_3}$, where the lattice parameter
along $z$ is larger than in the $xy$ plane and $Q_1$ is positive. 
On the other hand, the distortion $Q_2$ (opposite on the two 
sublattices of the $xy$-planes) corresponds to an alternate 
contraction and dilation of the Mn-O bonds on the $xy$ plane and 
mixes the $x$ and $z$
components of the $a,b$ orbitals
\begin{eqnarray}
|a\rangle & = & \cos(\theta/2) |x\rangle \pm \sin(\theta/2) |z\rangle
\nonumber \\
|b\rangle & = & \sin(\theta/2) |x\rangle \mp \cos(\theta/2) |z\rangle 
\end{eqnarray}
where $\tan(\theta)=Q_2/Q_1$ and 
the upper (lower) sign is for sites on sublattice 1 (2) of the
$xy$-planes. Accordingly the hoppings between the $a$ and $b$
orbitals of neighbouring sites can straightforwardly be obtained via Eqs. 
(\ref{xzhopping}). 
\begin{eqnarray}
t_{aa} & = & -t\left( 1 +2\cos\theta \right); \label{AA}\\
t_{ab} & = & -t\left(\pm \sqrt{3} +2\sin\theta \right); \;\;
t_{ba} =  -t\left(\mp \sqrt{3} +2\sin\theta \right); \label{ABBA}\\
t'_{aa} & = & 2t\left(1-\cos\theta \right); \;\;
t'_{ab}  =  t'_{ba}=-2t\sin\theta \label{AApABp}
\end{eqnarray}
the upper (lower) sign is for planar hopping in the $x$ ($y$) direction.

As customarily done, we assume that the Hund coupling 
between $\sigma=1/2$ $e_g$ electrons and the 
$S=3/2$ spin of the $t_{2g}$ electrons is 
so large that the initial and final states always have maximal
total spin $S_T=2$. Moreover, and most importantly, we also consider
a large local repulsion ($\sim U$) between electrons on Mn sites 
forbidding to two electrons to reside on the two $e_g$
levels of the same site. Then we work on a reduced
Hilbert space with only ${\rm Mn^{3+}}$ initial and final states
\cite{notaU}

We then carry out a perturbative calculation of both FM and AFM
magnetic couplings between sites 1 and 2 by considering second order
hopping processes from and to the ground state configuration
with one electron per Mn site occupying the lower $a$ orbital. We thus neglect 
the exchange-induced mixing of $e_g$ orbitals, considered by KK.
Notice that this last assumption relies on the JT splitting $\epsilon$
being substantially larger than both the temperature and the
superexchange scale $\sim t^2/U$.  
Due to the condition $S_T=2$, each site $i=1,2$ is five times degenerate,
$|2,m\rangle_i$ with $m=-2,-1,...,2$. The first step consists in
forming two-site states with given total spin $J=0,...,4$ from the
25 basis states $|2,m\rangle_1\otimes |2,m\rangle_2$. The suitable
Clebsch-Gordan coefficient are easily obtained. Within each $J$ 
subspace (the hopping processes conserve the total spin), the
hopping perturbation 
$$
H_t=-\sum_{\sigma;\alpha,\alpha'=a,b}t_{\alpha\alpha'} 
\left( c^\dagger_{1\sigma\alpha}c_{2\sigma\alpha}+
c^\dagger_{2\sigma\alpha}c_{1\sigma\alpha} \right)
$$
is applied twice to obtain the $\langle J,M | H_t^2 |J,M\rangle$
matrix elements. The double-hopping processes are of two types:
$aa$ and $ab$ depending on whether the initial $a$ electron hops
on the neighboring $a$ or $b$ orbitals.
Accordingly there are two superexchange channels, 
leading to couplings constants $J_{aa}$ and $J_{ab}$.
The easiest to be calculated is $J_{aa}$ since the Pauli principle 
forces the two initial $a$ electrons to have opposite spins. 
As a consequence only one intermediate virtual state per $J$ channel
is allowed, with one empty and one doubly occupied $a$ orbital.
Both the empty and the doubly occupied orbitals cost 
an energy $3J_H'/4$, where $J_H'$ is the Hund coupling between 
$e_g$ and $t_{2g}$ orbitals (the Hund energy is set to zero 
in the ${\rm Mn^{3+}}$ ground state configuration).
The doubly occupied $a$ orbital has an additional energy cost $U$.
All the intermediate states have then an energy $E_V=U+(3/2)J_H'$
above the ground state energy $E_0=0$. 
As it is standard, the perturbative energy gain can be compared with
the energies of $|J,M\rangle$ states 
as given by the effective Heisenberg model for the $aa$ channel
$$
H_{aa}=J_{aa}\left({\bf S}_1\cdot {\bf S}_2 +C\right)
$$
with $C$ being a constant energy shift to be determined.
The direct comparison provides, besides $C=-4$,
\begin{equation}
J_{aa}={1\over 4} {t_{aa}^2 \over U+(3/2)J_H'}
\label{jaa}
\end{equation} 
This is an effective AFM coupling between electrons on the
$xy$-plane. A similar expression is obtained for the interplane
coupling (i.e., in the $z$ direction) $J'_{aa}$, provided
$t_{aa}$ is replaced by $t_{aa}'$ in Eq. (\ref{AApABp}).

The calculation for $J_{ab}$ is slightly more complicated,
since the hopping electron can now give rise on the doubly
occupied site to both a $S_T=5/2$ or a $S_T=3/2$ state,
thus increasing the number of virtual states.
However, the same procedure illustrated above yields
$$
J_{ab}=-{{\overline{t_{ab}^2}} \over 40}
\left(
{8\over U'+\epsilon -J_H/2} \right. \nonumber
$$
\begin{equation}
-\left.
{3\over U'+\epsilon -J_H/2+5J_H'/2}-
{5\over U'+\epsilon +J_H/2+3J_H'/2}
\right)
\label{jab}
\end{equation}
where $U'$ is the local Coulomb repulsion 
between electrons on the $a$ and $b$ orbitals and 
\begin{equation}
{\overline{t_{ab}^2}}\equiv {(t_{ab}^2+t_{ba}^2) \over 2}=
t^2\left(3+4 \sin^2(\theta)\right).
\end{equation}
Again the analogous coupling in the $z$ direction can be 
obtained by replacing $t_{ab}$ and $t_{ba}$ with the corresponding
primed quantities of Eq. (\ref{AApABp}).
It can easily be seen that this coupling is ferromagnetic
and vanishes when $J_H$ and $J_H'$ are both zero.
Notice also that, since $J_{ab}$ arises from virtual hopping
$i\to j \to i$ and $j\to i \to j$ and since the $a$ and $b$
orbital combinations are reversed on neighboring sites,
the $\overline{t_{ab}^2}$ combination appears, which is the
same in the $x$ and $y$ directions [cf. instead (\ref{ABBA})]. Thus for each
crystalline direction one can write the effective Heisenberg couplings 
$J=J_{aa}+J_{ab}$
and $J'=J'_{aa}+J'_{ab}$ in the $xy$ planes and $z$ direction respectively.
Then the question arises concerning the parameter ranges 
such that the observed A-type AF is realized. In this case
the coupling must be dominantly ferromagnetic in the $xy$ planes and
dominantly AF in the $z$ direction. To this purpose we rewrite
the J's in the following more compact way
\begin{equation}
J_{aa}=t_{aa}^2/D_{aa}, \;\;
J_{ab}=-\overline{t_{ab}^2}/D_{ab}
\end{equation}
The condition that the $xy$ planes are ferromagnetically 
coupled is written as $|J_{ab}|>J_{aa}$, i.e.
\begin{equation}
\alpha^2 \equiv {\overline{t_{ab}^2} \over t_{aa}^2}> D.
\label{condxy}
\end{equation}
where $D={D_{ab} \over D_{aa}}$. 
At the same time, the condition for AF coupling 
in the $z$ direction
is expressed by $|J'_{ab}|<J'_{aa}$, that is
\begin{equation}
\alpha'^2 \equiv {{t'_{ab}}^2 \over {t'_{aa}}^2}< D.
\label{condz}
\end{equation}
Now, both $\alpha$ and $\alpha'$ (i.e. $t_{aa}$, $\overline{t}_{ab}$, 
$t'_{aa}$, and $t'_{ab}$) are funtions of $\theta$, or of the JT ratio
$Q_2/Q_1$.  Plotting $\alpha^2$ and $\alpha'^2$
as a function of $Q_2/|Q_1|$ for $Q_1<0$
(the relevant case for LMO), one obtains the curves in
Fig. 1. Since from the inequalities (\ref{condxy}) and (\ref{condz}),
one can deduce the condition
\begin{equation}
\alpha^2> D >\alpha'^2, 
\label{inequal}
\end{equation}
the value of $D$ should be below the solid curve and above the dashed ones.
The assumed orbital order becomes compatible with
A-AFMI order for  $Q_2/|Q_1| \gtrsim 2.6$, where the
relation (\ref{inequal}) can be satisfied. 
Notice instead that for $Q_1>0$ the curve $\alpha'^2$ 
is always smaller than $\alpha^2$, so 
that no restriction on the  $Q_2/Q_1$ ratio is needed to fulfill the
condition (\ref{inequal}). This is in agreement with the cooperation between 
superexchange and JT effect, found in this case, see for example the case of 
${\rm KCuF_3}$ \cite{KK}.
\begin{figure}
\begin{center}
% GNUPLOT: LaTeX picture
\setlength{\unitlength}{0.240900pt}
\ifx\plotpoint\undefined\newsavebox{\plotpoint}\fi
\sbox{\plotpoint}{\rule[-0.200pt]{0.400pt}{0.400pt}}%
\begin{picture}(1049,900)(0,0)
\font\gnuplot=cmr10 at 10pt
\gnuplot
\sbox{\plotpoint}{\rule[-0.200pt]{0.400pt}{0.400pt}}%
\put(220.0,113.0){\rule[-0.200pt]{184.288pt}{0.400pt}}
\put(220.0,113.0){\rule[-0.200pt]{0.400pt}{184.048pt}}
\put(220.0,113.0){\rule[-0.200pt]{4.818pt}{0.400pt}}
\put(198,113){\makebox(0,0)[r]{$0$}}
\put(965.0,113.0){\rule[-0.200pt]{4.818pt}{0.400pt}}
\put(220.0,209.0){\rule[-0.200pt]{4.818pt}{0.400pt}}
\put(198,209){\makebox(0,0)[r]{$1$}}
\put(965.0,209.0){\rule[-0.200pt]{4.818pt}{0.400pt}}
\put(220.0,304.0){\rule[-0.200pt]{4.818pt}{0.400pt}}
\put(198,304){\makebox(0,0)[r]{$2$}}
\put(965.0,304.0){\rule[-0.200pt]{4.818pt}{0.400pt}}
\put(220.0,400.0){\rule[-0.200pt]{4.818pt}{0.400pt}}
\put(198,400){\makebox(0,0)[r]{$3$}}
\put(965.0,400.0){\rule[-0.200pt]{4.818pt}{0.400pt}}
\put(220.0,495.0){\rule[-0.200pt]{4.818pt}{0.400pt}}
\put(198,495){\makebox(0,0)[r]{$4$}}
\put(965.0,495.0){\rule[-0.200pt]{4.818pt}{0.400pt}}
\put(220.0,591.0){\rule[-0.200pt]{4.818pt}{0.400pt}}
\put(198,591){\makebox(0,0)[r]{$5$}}
\put(965.0,591.0){\rule[-0.200pt]{4.818pt}{0.400pt}}
\put(220.0,686.0){\rule[-0.200pt]{4.818pt}{0.400pt}}
\put(198,686){\makebox(0,0)[r]{$6$}}
\put(965.0,686.0){\rule[-0.200pt]{4.818pt}{0.400pt}}
\put(220.0,782.0){\rule[-0.200pt]{4.818pt}{0.400pt}}
\put(198,782){\makebox(0,0)[r]{$7$}}
\put(965.0,782.0){\rule[-0.200pt]{4.818pt}{0.400pt}}
\put(220.0,877.0){\rule[-0.200pt]{4.818pt}{0.400pt}}
\put(198,877){\makebox(0,0)[r]{$8$}}
\put(965.0,877.0){\rule[-0.200pt]{4.818pt}{0.400pt}}
\put(220.0,113.0){\rule[-0.200pt]{0.400pt}{4.818pt}}
\put(220,68){\makebox(0,0){$0$}}
\put(220.0,857.0){\rule[-0.200pt]{0.400pt}{4.818pt}}
\put(329.0,113.0){\rule[-0.200pt]{0.400pt}{4.818pt}}
\put(329,68){\makebox(0,0){$1$}}
\put(329.0,857.0){\rule[-0.200pt]{0.400pt}{4.818pt}}
\put(439.0,113.0){\rule[-0.200pt]{0.400pt}{4.818pt}}
\put(439,68){\makebox(0,0){$2$}}
\put(439.0,857.0){\rule[-0.200pt]{0.400pt}{4.818pt}}
\put(548.0,113.0){\rule[-0.200pt]{0.400pt}{4.818pt}}
\put(548,68){\makebox(0,0){$3$}}
\put(548.0,857.0){\rule[-0.200pt]{0.400pt}{4.818pt}}
\put(657.0,113.0){\rule[-0.200pt]{0.400pt}{4.818pt}}
\put(657,68){\makebox(0,0){$4$}}
\put(657.0,857.0){\rule[-0.200pt]{0.400pt}{4.818pt}}
\put(766.0,113.0){\rule[-0.200pt]{0.400pt}{4.818pt}}
\put(766,68){\makebox(0,0){$5$}}
\put(766.0,857.0){\rule[-0.200pt]{0.400pt}{4.818pt}}
\put(876.0,113.0){\rule[-0.200pt]{0.400pt}{4.818pt}}
\put(876,68){\makebox(0,0){$6$}}
\put(876.0,857.0){\rule[-0.200pt]{0.400pt}{4.818pt}}
\put(985.0,113.0){\rule[-0.200pt]{0.400pt}{4.818pt}}
\put(985,68){\makebox(0,0){$7$}}
\put(985.0,857.0){\rule[-0.200pt]{0.400pt}{4.818pt}}
\put(220.0,113.0){\rule[-0.200pt]{184.288pt}{0.400pt}}
\put(985.0,113.0){\rule[-0.200pt]{0.400pt}{184.048pt}}
\put(220.0,877.0){\rule[-0.200pt]{184.288pt}{0.400pt}}
\put(45,495){\makebox(0,0){$\alpha^2 \,\,\, {\alpha^\prime}^2$}}
\put(602,23){\makebox(0,0){$\frac{Q_2}{|Q_1|}$}}
\put(220.0,113.0){\rule[-0.200pt]{0.400pt}{184.048pt}}
\sbox{\plotpoint}{\rule[-0.600pt]{1.200pt}{1.200pt}}%
\put(821,782){\makebox(0,0)[r]{$\alpha^2$}}
\put(843.0,782.0){\rule[-0.600pt]{15.899pt}{1.200pt}}
\put(220,145){\usebox{\plotpoint}}
\put(231,143.51){\rule{2.650pt}{1.200pt}}
\multiput(231.00,142.51)(5.500,2.000){2}{\rule{1.325pt}{1.200pt}}
\put(242,146.01){\rule{2.650pt}{1.200pt}}
\multiput(242.00,144.51)(5.500,3.000){2}{\rule{1.325pt}{1.200pt}}
\put(253,150.01){\rule{2.650pt}{1.200pt}}
\multiput(253.00,147.51)(5.500,5.000){2}{\rule{1.325pt}{1.200pt}}
\put(264,155.01){\rule{2.650pt}{1.200pt}}
\multiput(264.00,152.51)(5.500,5.000){2}{\rule{1.325pt}{1.200pt}}
\multiput(275.00,162.24)(0.622,0.509){2}{\rule{2.500pt}{0.123pt}}
\multiput(275.00,157.51)(5.811,6.000){2}{\rule{1.250pt}{1.200pt}}
\multiput(286.00,168.24)(0.622,0.509){2}{\rule{2.500pt}{0.123pt}}
\multiput(286.00,163.51)(5.811,6.000){2}{\rule{1.250pt}{1.200pt}}
\multiput(297.00,174.24)(0.546,0.505){4}{\rule{2.014pt}{0.122pt}}
\multiput(297.00,169.51)(5.819,7.000){2}{\rule{1.007pt}{1.200pt}}
\multiput(307.00,181.24)(0.581,0.503){6}{\rule{1.950pt}{0.121pt}}
\multiput(307.00,176.51)(6.953,8.000){2}{\rule{0.975pt}{1.200pt}}
\multiput(318.00,189.24)(0.581,0.503){6}{\rule{1.950pt}{0.121pt}}
\multiput(318.00,184.51)(6.953,8.000){2}{\rule{0.975pt}{1.200pt}}
\multiput(329.00,197.24)(0.581,0.503){6}{\rule{1.950pt}{0.121pt}}
\multiput(329.00,192.51)(6.953,8.000){2}{\rule{0.975pt}{1.200pt}}
\multiput(340.00,205.24)(0.581,0.503){6}{\rule{1.950pt}{0.121pt}}
\multiput(340.00,200.51)(6.953,8.000){2}{\rule{0.975pt}{1.200pt}}
\multiput(351.00,213.24)(0.524,0.502){8}{\rule{1.767pt}{0.121pt}}
\multiput(351.00,208.51)(7.333,9.000){2}{\rule{0.883pt}{1.200pt}}
\multiput(362.00,222.24)(0.581,0.503){6}{\rule{1.950pt}{0.121pt}}
\multiput(362.00,217.51)(6.953,8.000){2}{\rule{0.975pt}{1.200pt}}
\multiput(373.00,230.24)(0.524,0.502){8}{\rule{1.767pt}{0.121pt}}
\multiput(373.00,225.51)(7.333,9.000){2}{\rule{0.883pt}{1.200pt}}
\multiput(384.00,239.24)(0.581,0.503){6}{\rule{1.950pt}{0.121pt}}
\multiput(384.00,234.51)(6.953,8.000){2}{\rule{0.975pt}{1.200pt}}
\multiput(395.00,247.24)(0.524,0.502){8}{\rule{1.767pt}{0.121pt}}
\multiput(395.00,242.51)(7.333,9.000){2}{\rule{0.883pt}{1.200pt}}
\multiput(406.00,256.24)(0.581,0.503){6}{\rule{1.950pt}{0.121pt}}
\multiput(406.00,251.51)(6.953,8.000){2}{\rule{0.975pt}{1.200pt}}
\multiput(417.00,264.24)(0.581,0.503){6}{\rule{1.950pt}{0.121pt}}
\multiput(417.00,259.51)(6.953,8.000){2}{\rule{0.975pt}{1.200pt}}
\multiput(428.00,272.24)(0.581,0.503){6}{\rule{1.950pt}{0.121pt}}
\multiput(428.00,267.51)(6.953,8.000){2}{\rule{0.975pt}{1.200pt}}
\multiput(439.00,280.24)(0.581,0.503){6}{\rule{1.950pt}{0.121pt}}
\multiput(439.00,275.51)(6.953,8.000){2}{\rule{0.975pt}{1.200pt}}
\multiput(450.00,288.24)(0.506,0.503){6}{\rule{1.800pt}{0.121pt}}
\multiput(450.00,283.51)(6.264,8.000){2}{\rule{0.900pt}{1.200pt}}
\multiput(460.00,296.24)(0.642,0.505){4}{\rule{2.186pt}{0.122pt}}
\multiput(460.00,291.51)(6.463,7.000){2}{\rule{1.093pt}{1.200pt}}
\multiput(471.00,303.24)(0.581,0.503){6}{\rule{1.950pt}{0.121pt}}
\multiput(471.00,298.51)(6.953,8.000){2}{\rule{0.975pt}{1.200pt}}
\multiput(482.00,311.24)(0.642,0.505){4}{\rule{2.186pt}{0.122pt}}
\multiput(482.00,306.51)(6.463,7.000){2}{\rule{1.093pt}{1.200pt}}
\multiput(493.00,318.24)(0.642,0.505){4}{\rule{2.186pt}{0.122pt}}
\multiput(493.00,313.51)(6.463,7.000){2}{\rule{1.093pt}{1.200pt}}
\multiput(504.00,325.24)(0.642,0.505){4}{\rule{2.186pt}{0.122pt}}
\multiput(504.00,320.51)(6.463,7.000){2}{\rule{1.093pt}{1.200pt}}
\multiput(515.00,332.24)(0.622,0.509){2}{\rule{2.500pt}{0.123pt}}
\multiput(515.00,327.51)(5.811,6.000){2}{\rule{1.250pt}{1.200pt}}
\multiput(526.00,338.24)(0.642,0.505){4}{\rule{2.186pt}{0.122pt}}
\multiput(526.00,333.51)(6.463,7.000){2}{\rule{1.093pt}{1.200pt}}
\multiput(537.00,345.24)(0.642,0.505){4}{\rule{2.186pt}{0.122pt}}
\multiput(537.00,340.51)(6.463,7.000){2}{\rule{1.093pt}{1.200pt}}
\multiput(548.00,352.24)(0.622,0.509){2}{\rule{2.500pt}{0.123pt}}
\multiput(548.00,347.51)(5.811,6.000){2}{\rule{1.250pt}{1.200pt}}
\multiput(559.00,358.24)(0.622,0.509){2}{\rule{2.500pt}{0.123pt}}
\multiput(559.00,353.51)(5.811,6.000){2}{\rule{1.250pt}{1.200pt}}
\multiput(570.00,364.24)(0.622,0.509){2}{\rule{2.500pt}{0.123pt}}
\multiput(570.00,359.51)(5.811,6.000){2}{\rule{1.250pt}{1.200pt}}
\multiput(581.00,370.24)(0.622,0.509){2}{\rule{2.500pt}{0.123pt}}
\multiput(581.00,365.51)(5.811,6.000){2}{\rule{1.250pt}{1.200pt}}
\put(592,374.01){\rule{2.650pt}{1.200pt}}
\multiput(592.00,371.51)(5.500,5.000){2}{\rule{1.325pt}{1.200pt}}
\multiput(603.00,381.24)(0.452,0.509){2}{\rule{2.300pt}{0.123pt}}
\multiput(603.00,376.51)(5.226,6.000){2}{\rule{1.150pt}{1.200pt}}
\put(613,385.01){\rule{2.650pt}{1.200pt}}
\multiput(613.00,382.51)(5.500,5.000){2}{\rule{1.325pt}{1.200pt}}
\multiput(624.00,392.24)(0.622,0.509){2}{\rule{2.500pt}{0.123pt}}
\multiput(624.00,387.51)(5.811,6.000){2}{\rule{1.250pt}{1.200pt}}
\put(635,396.01){\rule{2.650pt}{1.200pt}}
\multiput(635.00,393.51)(5.500,5.000){2}{\rule{1.325pt}{1.200pt}}
\put(646,401.01){\rule{2.650pt}{1.200pt}}
\multiput(646.00,398.51)(5.500,5.000){2}{\rule{1.325pt}{1.200pt}}
\put(657,406.01){\rule{2.650pt}{1.200pt}}
\multiput(657.00,403.51)(5.500,5.000){2}{\rule{1.325pt}{1.200pt}}
\put(668,411.01){\rule{2.650pt}{1.200pt}}
\multiput(668.00,408.51)(5.500,5.000){2}{\rule{1.325pt}{1.200pt}}
\put(679,415.51){\rule{2.650pt}{1.200pt}}
\multiput(679.00,413.51)(5.500,4.000){2}{\rule{1.325pt}{1.200pt}}
\put(690,420.01){\rule{2.650pt}{1.200pt}}
\multiput(690.00,417.51)(5.500,5.000){2}{\rule{1.325pt}{1.200pt}}
\put(701,424.51){\rule{2.650pt}{1.200pt}}
\multiput(701.00,422.51)(5.500,4.000){2}{\rule{1.325pt}{1.200pt}}
\put(712,429.01){\rule{2.650pt}{1.200pt}}
\multiput(712.00,426.51)(5.500,5.000){2}{\rule{1.325pt}{1.200pt}}
\put(723,433.51){\rule{2.650pt}{1.200pt}}
\multiput(723.00,431.51)(5.500,4.000){2}{\rule{1.325pt}{1.200pt}}
\put(734,437.51){\rule{2.650pt}{1.200pt}}
\multiput(734.00,435.51)(5.500,4.000){2}{\rule{1.325pt}{1.200pt}}
\put(745,441.51){\rule{2.650pt}{1.200pt}}
\multiput(745.00,439.51)(5.500,4.000){2}{\rule{1.325pt}{1.200pt}}
\put(756,445.51){\rule{2.409pt}{1.200pt}}
\multiput(756.00,443.51)(5.000,4.000){2}{\rule{1.204pt}{1.200pt}}
\put(766,449.51){\rule{2.650pt}{1.200pt}}
\multiput(766.00,447.51)(5.500,4.000){2}{\rule{1.325pt}{1.200pt}}
\put(777,453.51){\rule{2.650pt}{1.200pt}}
\multiput(777.00,451.51)(5.500,4.000){2}{\rule{1.325pt}{1.200pt}}
\put(788,457.51){\rule{2.650pt}{1.200pt}}
\multiput(788.00,455.51)(5.500,4.000){2}{\rule{1.325pt}{1.200pt}}
\put(799,461.01){\rule{2.650pt}{1.200pt}}
\multiput(799.00,459.51)(5.500,3.000){2}{\rule{1.325pt}{1.200pt}}
\put(810,464.51){\rule{2.650pt}{1.200pt}}
\multiput(810.00,462.51)(5.500,4.000){2}{\rule{1.325pt}{1.200pt}}
\put(821,468.01){\rule{2.650pt}{1.200pt}}
\multiput(821.00,466.51)(5.500,3.000){2}{\rule{1.325pt}{1.200pt}}
\put(832,471.51){\rule{2.650pt}{1.200pt}}
\multiput(832.00,469.51)(5.500,4.000){2}{\rule{1.325pt}{1.200pt}}
\put(843,475.01){\rule{2.650pt}{1.200pt}}
\multiput(843.00,473.51)(5.500,3.000){2}{\rule{1.325pt}{1.200pt}}
\put(854,478.51){\rule{2.650pt}{1.200pt}}
\multiput(854.00,476.51)(5.500,4.000){2}{\rule{1.325pt}{1.200pt}}
\put(865,482.01){\rule{2.650pt}{1.200pt}}
\multiput(865.00,480.51)(5.500,3.000){2}{\rule{1.325pt}{1.200pt}}
\put(876,485.01){\rule{2.650pt}{1.200pt}}
\multiput(876.00,483.51)(5.500,3.000){2}{\rule{1.325pt}{1.200pt}}
\put(887,488.01){\rule{2.650pt}{1.200pt}}
\multiput(887.00,486.51)(5.500,3.000){2}{\rule{1.325pt}{1.200pt}}
\put(898,491.01){\rule{2.409pt}{1.200pt}}
\multiput(898.00,489.51)(5.000,3.000){2}{\rule{1.204pt}{1.200pt}}
\put(908,494.01){\rule{2.650pt}{1.200pt}}
\multiput(908.00,492.51)(5.500,3.000){2}{\rule{1.325pt}{1.200pt}}
\put(919,497.01){\rule{2.650pt}{1.200pt}}
\multiput(919.00,495.51)(5.500,3.000){2}{\rule{1.325pt}{1.200pt}}
\put(930,500.01){\rule{2.650pt}{1.200pt}}
\multiput(930.00,498.51)(5.500,3.000){2}{\rule{1.325pt}{1.200pt}}
\put(941,503.01){\rule{2.650pt}{1.200pt}}
\multiput(941.00,501.51)(5.500,3.000){2}{\rule{1.325pt}{1.200pt}}
\put(952,505.51){\rule{2.650pt}{1.200pt}}
\multiput(952.00,504.51)(5.500,2.000){2}{\rule{1.325pt}{1.200pt}}
\put(963,508.01){\rule{2.650pt}{1.200pt}}
\multiput(963.00,506.51)(5.500,3.000){2}{\rule{1.325pt}{1.200pt}}
\put(220.0,145.0){\rule[-0.600pt]{2.650pt}{1.200pt}}
\sbox{\plotpoint}{\rule[-0.400pt]{0.800pt}{0.800pt}}%
\put(821,737){\makebox(0,0)[r]{${\alpha^\prime}^2$}}
\put(843.0,737.0){\rule[-0.400pt]{15.899pt}{0.800pt}}
\put(220,209){\usebox{\plotpoint}}
\multiput(309.40,836.65)(0.514,-6.554){13}{\rule{0.124pt}{9.720pt}}
\multiput(306.34,856.83)(10.000,-98.826){2}{\rule{0.800pt}{4.860pt}}
\multiput(319.40,729.70)(0.512,-4.478){15}{\rule{0.123pt}{6.818pt}}
\multiput(316.34,743.85)(11.000,-76.849){2}{\rule{0.800pt}{3.409pt}}
\multiput(330.40,645.64)(0.512,-3.331){15}{\rule{0.123pt}{5.145pt}}
\multiput(327.34,656.32)(11.000,-57.320){2}{\rule{0.800pt}{2.573pt}}
\multiput(341.40,582.47)(0.512,-2.534){15}{\rule{0.123pt}{3.982pt}}
\multiput(338.34,590.74)(11.000,-43.736){2}{\rule{0.800pt}{1.991pt}}
\multiput(352.40,533.49)(0.512,-2.035){15}{\rule{0.123pt}{3.255pt}}
\multiput(349.34,540.25)(11.000,-35.245){2}{\rule{0.800pt}{1.627pt}}
\multiput(363.40,494.21)(0.512,-1.586){15}{\rule{0.123pt}{2.600pt}}
\multiput(360.34,499.60)(11.000,-27.604){2}{\rule{0.800pt}{1.300pt}}
\multiput(374.40,463.02)(0.512,-1.287){15}{\rule{0.123pt}{2.164pt}}
\multiput(371.34,467.51)(11.000,-22.509){2}{\rule{0.800pt}{1.082pt}}
\multiput(385.40,437.53)(0.512,-1.038){15}{\rule{0.123pt}{1.800pt}}
\multiput(382.34,441.26)(11.000,-18.264){2}{\rule{0.800pt}{0.900pt}}
\multiput(396.40,416.43)(0.512,-0.888){15}{\rule{0.123pt}{1.582pt}}
\multiput(393.34,419.72)(11.000,-15.717){2}{\rule{0.800pt}{0.791pt}}
\multiput(407.40,398.34)(0.512,-0.739){15}{\rule{0.123pt}{1.364pt}}
\multiput(404.34,401.17)(11.000,-13.170){2}{\rule{0.800pt}{0.682pt}}
\multiput(418.40,382.94)(0.512,-0.639){15}{\rule{0.123pt}{1.218pt}}
\multiput(415.34,385.47)(11.000,-11.472){2}{\rule{0.800pt}{0.609pt}}
\multiput(429.40,369.55)(0.512,-0.539){15}{\rule{0.123pt}{1.073pt}}
\multiput(426.34,371.77)(11.000,-9.774){2}{\rule{0.800pt}{0.536pt}}
\multiput(439.00,360.08)(0.543,-0.514){13}{\rule{1.080pt}{0.124pt}}
\multiput(439.00,360.34)(8.758,-10.000){2}{\rule{0.540pt}{0.800pt}}
\multiput(450.00,350.08)(0.548,-0.516){11}{\rule{1.089pt}{0.124pt}}
\multiput(450.00,350.34)(7.740,-9.000){2}{\rule{0.544pt}{0.800pt}}
\multiput(460.00,341.08)(0.700,-0.520){9}{\rule{1.300pt}{0.125pt}}
\multiput(460.00,341.34)(8.302,-8.000){2}{\rule{0.650pt}{0.800pt}}
\multiput(471.00,333.08)(0.700,-0.520){9}{\rule{1.300pt}{0.125pt}}
\multiput(471.00,333.34)(8.302,-8.000){2}{\rule{0.650pt}{0.800pt}}
\multiput(482.00,325.07)(1.020,-0.536){5}{\rule{1.667pt}{0.129pt}}
\multiput(482.00,325.34)(7.541,-6.000){2}{\rule{0.833pt}{0.800pt}}
\multiput(493.00,319.07)(1.020,-0.536){5}{\rule{1.667pt}{0.129pt}}
\multiput(493.00,319.34)(7.541,-6.000){2}{\rule{0.833pt}{0.800pt}}
\multiput(504.00,313.06)(1.432,-0.560){3}{\rule{1.960pt}{0.135pt}}
\multiput(504.00,313.34)(6.932,-5.000){2}{\rule{0.980pt}{0.800pt}}
\multiput(515.00,308.06)(1.432,-0.560){3}{\rule{1.960pt}{0.135pt}}
\multiput(515.00,308.34)(6.932,-5.000){2}{\rule{0.980pt}{0.800pt}}
\multiput(526.00,303.06)(1.432,-0.560){3}{\rule{1.960pt}{0.135pt}}
\multiput(526.00,303.34)(6.932,-5.000){2}{\rule{0.980pt}{0.800pt}}
\put(537,296.34){\rule{2.400pt}{0.800pt}}
\multiput(537.00,298.34)(6.019,-4.000){2}{\rule{1.200pt}{0.800pt}}
\put(548,292.84){\rule{2.650pt}{0.800pt}}
\multiput(548.00,294.34)(5.500,-3.000){2}{\rule{1.325pt}{0.800pt}}
\put(559,289.34){\rule{2.400pt}{0.800pt}}
\multiput(559.00,291.34)(6.019,-4.000){2}{\rule{1.200pt}{0.800pt}}
\put(570,285.84){\rule{2.650pt}{0.800pt}}
\multiput(570.00,287.34)(5.500,-3.000){2}{\rule{1.325pt}{0.800pt}}
\put(581,282.84){\rule{2.650pt}{0.800pt}}
\multiput(581.00,284.34)(5.500,-3.000){2}{\rule{1.325pt}{0.800pt}}
\put(592,279.84){\rule{2.650pt}{0.800pt}}
\multiput(592.00,281.34)(5.500,-3.000){2}{\rule{1.325pt}{0.800pt}}
\put(603,277.34){\rule{2.409pt}{0.800pt}}
\multiput(603.00,278.34)(5.000,-2.000){2}{\rule{1.204pt}{0.800pt}}
\put(613,275.34){\rule{2.650pt}{0.800pt}}
\multiput(613.00,276.34)(5.500,-2.000){2}{\rule{1.325pt}{0.800pt}}
\put(624,272.84){\rule{2.650pt}{0.800pt}}
\multiput(624.00,274.34)(5.500,-3.000){2}{\rule{1.325pt}{0.800pt}}
\put(635,270.34){\rule{2.650pt}{0.800pt}}
\multiput(635.00,271.34)(5.500,-2.000){2}{\rule{1.325pt}{0.800pt}}
\put(646,268.34){\rule{2.650pt}{0.800pt}}
\multiput(646.00,269.34)(5.500,-2.000){2}{\rule{1.325pt}{0.800pt}}
\put(657,266.34){\rule{2.650pt}{0.800pt}}
\multiput(657.00,267.34)(5.500,-2.000){2}{\rule{1.325pt}{0.800pt}}
\put(668,264.84){\rule{2.650pt}{0.800pt}}
\multiput(668.00,265.34)(5.500,-1.000){2}{\rule{1.325pt}{0.800pt}}
\put(679,263.34){\rule{2.650pt}{0.800pt}}
\multiput(679.00,264.34)(5.500,-2.000){2}{\rule{1.325pt}{0.800pt}}
\put(690,261.84){\rule{2.650pt}{0.800pt}}
\multiput(690.00,262.34)(5.500,-1.000){2}{\rule{1.325pt}{0.800pt}}
\put(701,260.34){\rule{2.650pt}{0.800pt}}
\multiput(701.00,261.34)(5.500,-2.000){2}{\rule{1.325pt}{0.800pt}}
\put(712,258.84){\rule{2.650pt}{0.800pt}}
\multiput(712.00,259.34)(5.500,-1.000){2}{\rule{1.325pt}{0.800pt}}
\put(723,257.34){\rule{2.650pt}{0.800pt}}
\multiput(723.00,258.34)(5.500,-2.000){2}{\rule{1.325pt}{0.800pt}}
\put(734,255.84){\rule{2.650pt}{0.800pt}}
\multiput(734.00,256.34)(5.500,-1.000){2}{\rule{1.325pt}{0.800pt}}
\put(745,254.84){\rule{2.650pt}{0.800pt}}
\multiput(745.00,255.34)(5.500,-1.000){2}{\rule{1.325pt}{0.800pt}}
\put(756,253.84){\rule{2.409pt}{0.800pt}}
\multiput(756.00,254.34)(5.000,-1.000){2}{\rule{1.204pt}{0.800pt}}
\put(766,252.84){\rule{2.650pt}{0.800pt}}
\multiput(766.00,253.34)(5.500,-1.000){2}{\rule{1.325pt}{0.800pt}}
\put(777,251.84){\rule{2.650pt}{0.800pt}}
\multiput(777.00,252.34)(5.500,-1.000){2}{\rule{1.325pt}{0.800pt}}
\put(788,250.84){\rule{2.650pt}{0.800pt}}
\multiput(788.00,251.34)(5.500,-1.000){2}{\rule{1.325pt}{0.800pt}}
\put(799,249.84){\rule{2.650pt}{0.800pt}}
\multiput(799.00,250.34)(5.500,-1.000){2}{\rule{1.325pt}{0.800pt}}
\put(810,248.84){\rule{2.650pt}{0.800pt}}
\multiput(810.00,249.34)(5.500,-1.000){2}{\rule{1.325pt}{0.800pt}}
\put(821,247.84){\rule{2.650pt}{0.800pt}}
\multiput(821.00,248.34)(5.500,-1.000){2}{\rule{1.325pt}{0.800pt}}
\put(832,246.84){\rule{2.650pt}{0.800pt}}
\multiput(832.00,247.34)(5.500,-1.000){2}{\rule{1.325pt}{0.800pt}}
\put(843,245.84){\rule{2.650pt}{0.800pt}}
\multiput(843.00,246.34)(5.500,-1.000){2}{\rule{1.325pt}{0.800pt}}
\put(220.0,209.0){\rule[-0.400pt]{0.800pt}{160.921pt}}
\put(865,244.84){\rule{2.650pt}{0.800pt}}
\multiput(865.00,245.34)(5.500,-1.000){2}{\rule{1.325pt}{0.800pt}}
\put(876,243.84){\rule{2.650pt}{0.800pt}}
\multiput(876.00,244.34)(5.500,-1.000){2}{\rule{1.325pt}{0.800pt}}
\put(887,242.84){\rule{2.650pt}{0.800pt}}
\multiput(887.00,243.34)(5.500,-1.000){2}{\rule{1.325pt}{0.800pt}}
\put(854.0,247.0){\rule[-0.400pt]{2.650pt}{0.800pt}}
\put(908,241.84){\rule{2.650pt}{0.800pt}}
\multiput(908.00,242.34)(5.500,-1.000){2}{\rule{1.325pt}{0.800pt}}
\put(919,240.84){\rule{2.650pt}{0.800pt}}
\multiput(919.00,241.34)(5.500,-1.000){2}{\rule{1.325pt}{0.800pt}}
\put(898.0,244.0){\rule[-0.400pt]{2.409pt}{0.800pt}}
\put(941,239.84){\rule{2.650pt}{0.800pt}}
\multiput(941.00,240.34)(5.500,-1.000){2}{\rule{1.325pt}{0.800pt}}
\put(930.0,242.0){\rule[-0.400pt]{2.650pt}{0.800pt}}
\put(963,238.84){\rule{2.650pt}{0.800pt}}
\multiput(963.00,239.34)(5.500,-1.000){2}{\rule{1.325pt}{0.800pt}}
\put(952.0,241.0){\rule[-0.400pt]{2.650pt}{0.800pt}}
\sbox{\plotpoint}{\rule[-0.200pt]{0.400pt}{0.400pt}}%
\put(821,692){\makebox(0,0)[r]{${\alpha^\prime}^2_b$}}
\multiput(843,692)(20.756,0.000){4}{\usebox{\plotpoint}}
\put(909,692){\usebox{\plotpoint}}
\multiput(293,877)(1.562,-20.697){3}{\usebox{\plotpoint}}
\multiput(297,824)(1.522,-20.700){3}{\usebox{\plotpoint}}
\multiput(302,756)(1.814,-20.676){3}{\usebox{\plotpoint}}
\multiput(307,699)(2.574,-20.595){2}{\usebox{\plotpoint}}
\multiput(313,651)(2.574,-20.595){2}{\usebox{\plotpoint}}
\multiput(318,611)(3.507,-20.457){2}{\usebox{\plotpoint}}
\multiput(324,576)(3.305,-20.491){2}{\usebox{\plotpoint}}
\put(333.20,526.78){\usebox{\plotpoint}}
\put(337.60,506.50){\usebox{\plotpoint}}
\put(342.59,486.37){\usebox{\plotpoint}}
\put(348.38,466.44){\usebox{\plotpoint}}
\put(354.65,446.67){\usebox{\plotpoint}}
\put(361.60,427.11){\usebox{\plotpoint}}
\multiput(362,426)(8.698,-18.845){0}{\usebox{\plotpoint}}
\put(370.02,408.15){\usebox{\plotpoint}}
\multiput(373,401)(8.589,-18.895){0}{\usebox{\plotpoint}}
\put(378.48,389.20){\usebox{\plotpoint}}
\put(388.87,371.24){\usebox{\plotpoint}}
\multiput(389,371)(12.453,-16.604){0}{\usebox{\plotpoint}}
\multiput(395,363)(11.000,-17.601){0}{\usebox{\plotpoint}}
\put(400.68,354.20){\usebox{\plotpoint}}
\multiput(406,348)(13.287,-15.945){0}{\usebox{\plotpoint}}
\put(414.37,338.63){\usebox{\plotpoint}}
\multiput(417,336)(13.287,-15.945){0}{\usebox{\plotpoint}}
\multiput(422,330)(15.945,-13.287){0}{\usebox{\plotpoint}}
\put(429.01,323.99){\usebox{\plotpoint}}
\multiput(433,320)(17.270,-11.513){0}{\usebox{\plotpoint}}
\multiput(439,316)(16.207,-12.966){0}{\usebox{\plotpoint}}
\put(445.24,311.17){\usebox{\plotpoint}}
\multiput(450,308)(16.207,-12.966){0}{\usebox{\plotpoint}}
\multiput(455,304)(16.207,-12.966){0}{\usebox{\plotpoint}}
\put(462.00,299.00){\usebox{\plotpoint}}
\multiput(466,297)(17.798,-10.679){0}{\usebox{\plotpoint}}
\multiput(471,294)(18.564,-9.282){0}{\usebox{\plotpoint}}
\put(480.21,289.08){\usebox{\plotpoint}}
\multiput(482,288)(19.690,-6.563){0}{\usebox{\plotpoint}}
\multiput(488,286)(17.798,-10.679){0}{\usebox{\plotpoint}}
\multiput(493,283)(19.690,-6.563){0}{\usebox{\plotpoint}}
\put(499.16,280.91){\usebox{\plotpoint}}
\multiput(504,278)(19.690,-6.563){0}{\usebox{\plotpoint}}
\multiput(510,276)(19.271,-7.708){0}{\usebox{\plotpoint}}
\put(518.22,272.93){\usebox{\plotpoint}}
\multiput(521,272)(19.271,-7.708){0}{\usebox{\plotpoint}}
\multiput(526,270)(20.352,-4.070){0}{\usebox{\plotpoint}}
\multiput(531,269)(19.690,-6.563){0}{\usebox{\plotpoint}}
\put(537.95,266.62){\usebox{\plotpoint}}
\multiput(542,265)(20.473,-3.412){0}{\usebox{\plotpoint}}
\multiput(548,264)(19.271,-7.708){0}{\usebox{\plotpoint}}
\put(557.86,261.19){\usebox{\plotpoint}}
\multiput(559,261)(19.271,-7.708){0}{\usebox{\plotpoint}}
\multiput(564,259)(20.473,-3.412){0}{\usebox{\plotpoint}}
\multiput(570,258)(20.352,-4.070){0}{\usebox{\plotpoint}}
\put(577.87,256.04){\usebox{\plotpoint}}
\multiput(581,255)(20.352,-4.070){0}{\usebox{\plotpoint}}
\multiput(586,254)(20.473,-3.412){0}{\usebox{\plotpoint}}
\multiput(592,253)(20.352,-4.070){0}{\usebox{\plotpoint}}
\put(598.16,251.81){\usebox{\plotpoint}}
\multiput(603,251)(20.352,-4.070){0}{\usebox{\plotpoint}}
\multiput(608,250)(20.352,-4.070){0}{\usebox{\plotpoint}}
\put(618.58,248.07){\usebox{\plotpoint}}
\multiput(619,248)(20.352,-4.070){0}{\usebox{\plotpoint}}
\multiput(624,247)(20.473,-3.412){0}{\usebox{\plotpoint}}
\multiput(630,246)(20.352,-4.070){0}{\usebox{\plotpoint}}
\put(638.99,244.33){\usebox{\plotpoint}}
\multiput(641,244)(20.352,-4.070){0}{\usebox{\plotpoint}}
\multiput(646,243)(20.473,-3.412){0}{\usebox{\plotpoint}}
\multiput(652,242)(20.756,0.000){0}{\usebox{\plotpoint}}
\put(659.50,241.58){\usebox{\plotpoint}}
\multiput(663,241)(20.352,-4.070){0}{\usebox{\plotpoint}}
\multiput(668,240)(20.756,0.000){0}{\usebox{\plotpoint}}
\multiput(674,240)(20.352,-4.070){0}{\usebox{\plotpoint}}
\put(679.99,238.80){\usebox{\plotpoint}}
\multiput(684,238)(20.473,-3.412){0}{\usebox{\plotpoint}}
\multiput(690,237)(20.756,0.000){0}{\usebox{\plotpoint}}
\put(700.51,236.08){\usebox{\plotpoint}}
\multiput(701,236)(20.756,0.000){0}{\usebox{\plotpoint}}
\multiput(706,236)(20.473,-3.412){0}{\usebox{\plotpoint}}
\multiput(712,235)(20.352,-4.070){0}{\usebox{\plotpoint}}
\put(721.08,234.00){\usebox{\plotpoint}}
\multiput(723,234)(20.352,-4.070){0}{\usebox{\plotpoint}}
\multiput(728,233)(20.756,0.000){0}{\usebox{\plotpoint}}
\multiput(734,233)(20.352,-4.070){0}{\usebox{\plotpoint}}
\put(741.63,232.00){\usebox{\plotpoint}}
\multiput(745,232)(20.352,-4.070){0}{\usebox{\plotpoint}}
\multiput(750,231)(20.756,0.000){0}{\usebox{\plotpoint}}
\multiput(755,231)(20.473,-3.412){0}{\usebox{\plotpoint}}
\put(762.21,230.00){\usebox{\plotpoint}}
\multiput(766,230)(20.473,-3.412){0}{\usebox{\plotpoint}}
\multiput(772,229)(20.756,0.000){0}{\usebox{\plotpoint}}
\put(782.88,229.00){\usebox{\plotpoint}}
\multiput(783,229)(20.352,-4.070){0}{\usebox{\plotpoint}}
\multiput(788,228)(20.756,0.000){0}{\usebox{\plotpoint}}
\multiput(794,228)(20.352,-4.070){0}{\usebox{\plotpoint}}
\put(803.44,227.00){\usebox{\plotpoint}}
\multiput(805,227)(20.756,0.000){0}{\usebox{\plotpoint}}
\multiput(810,227)(20.473,-3.412){0}{\usebox{\plotpoint}}
\multiput(816,226)(20.756,0.000){0}{\usebox{\plotpoint}}
\put(824.07,225.49){\usebox{\plotpoint}}
\multiput(827,225)(20.756,0.000){0}{\usebox{\plotpoint}}
\multiput(832,225)(20.756,0.000){0}{\usebox{\plotpoint}}
\multiput(837,225)(20.473,-3.412){0}{\usebox{\plotpoint}}
\put(844.70,224.00){\usebox{\plotpoint}}
\multiput(848,224)(20.756,0.000){0}{\usebox{\plotpoint}}
\multiput(854,224)(20.352,-4.070){0}{\usebox{\plotpoint}}
\multiput(859,223)(20.756,0.000){0}{\usebox{\plotpoint}}
\put(865.36,223.00){\usebox{\plotpoint}}
\multiput(870,223)(20.756,0.000){0}{\usebox{\plotpoint}}
\multiput(876,223)(20.352,-4.070){0}{\usebox{\plotpoint}}
\put(886.01,222.00){\usebox{\plotpoint}}
\multiput(887,222)(20.756,0.000){0}{\usebox{\plotpoint}}
\multiput(892,222)(20.473,-3.412){0}{\usebox{\plotpoint}}
\multiput(898,221)(20.756,0.000){0}{\usebox{\plotpoint}}
\put(906.69,221.00){\usebox{\plotpoint}}
\multiput(908,221)(20.756,0.000){0}{\usebox{\plotpoint}}
\multiput(914,221)(20.352,-4.070){0}{\usebox{\plotpoint}}
\multiput(919,220)(20.756,0.000){0}{\usebox{\plotpoint}}
\put(927.34,220.00){\usebox{\plotpoint}}
\multiput(930,220)(20.756,0.000){0}{\usebox{\plotpoint}}
\multiput(936,220)(20.352,-4.070){0}{\usebox{\plotpoint}}
\multiput(941,219)(20.756,0.000){0}{\usebox{\plotpoint}}
\put(948.00,219.00){\usebox{\plotpoint}}
\multiput(952,219)(20.756,0.000){0}{\usebox{\plotpoint}}
\multiput(958,219)(20.352,-4.070){0}{\usebox{\plotpoint}}
\put(968.66,218.00){\usebox{\plotpoint}}
\multiput(969,218)(20.756,0.000){0}{\usebox{\plotpoint}}
\multiput(974,218)(20.756,0.000){0}{\usebox{\plotpoint}}
\multiput(980,218)(20.756,0.000){0}{\usebox{\plotpoint}}
\put(985,218){\usebox{\plotpoint}}
\sbox{\plotpoint}{\rule[-0.500pt]{1.000pt}{1.000pt}}%
\put(821,647){\makebox(0,0)[r]{${Q_2}/{|Q_1|}=2.84$}}
\multiput(843,647)(20.756,0.000){4}{\usebox{\plotpoint}}
\put(909,647){\usebox{\plotpoint}}
\put(530,113){\usebox{\plotpoint}}
\multiput(530,113)(0.000,20.756){37}{\usebox{\plotpoint}}
\put(530,877){\usebox{\plotpoint}}
\put(821,602){\makebox(0,0)[r]{${Q_2}/{|Q_1|}=6.69$}}
\multiput(843,602)(20.756,0.000){4}{\usebox{\plotpoint}}
\put(909,602){\usebox{\plotpoint}}
\put(951,113){\usebox{\plotpoint}}
\multiput(951,113)(0.000,20.756){37}{\usebox{\plotpoint}}
\put(951,877){\usebox{\plotpoint}}
\end{picture}
\end{center}
{\small Fig. 1.: $\alpha^2$ (solid line) and $\alpha'^2$ (dashed line)
as a function of the JT ratio $Q_2/|Q_1|$ for negative $Q_1$.
The dotted line represents $\alpha'^2$ with $t'_{xz}=0.05t$.
The vertical lines indicate the experimental values $Q_2/|Q_1|=2.84,6.69$
(see text)}
\end{figure}
More precisely we find that
\begin{equation}
\frac{J}{J'}=\frac{t_{aa}^2}{{t'_{aa}}^2} 
\frac{D-\alpha^2}{D-\alpha'^2}
\label{JJ}
\end{equation}

Notice that a substantial amount of $Q_2/|Q_1|$ JT
distortion is needed to leave the possility open
for the condition (\ref{inequal}) to be fulfilled.
Using standard results \cite{goodbook} connecting the structural
parameters with the $Q_2/|Q_1|$ ratio ($Q_2/|Q_3|$ in the notation of
Ref. \cite{goodbook}) we estimated $Q_2/|Q_1|\approx 1.97$ and $6.69$
for the parameters of the orthorhombic structures reported in 
Table V of Ref.\cite{mitchell}, while a value 2.84 is obtained from the data
related to the crystal where the abovementioned values of $J_{AF}$ and $J_F$ 
have been measured \cite{moussa}. Although not all  values
are compatible with the condition $Q_2/|Q_1| \gtrsim 2.6$,
the large variability of the resulting estimates for $Q_2/|Q_1|$ 
 indicates that
this parameter may vary from a crystal to another so that
the condition $Q_2/|Q_1|\gtrsim 2.6$ is quite reasonable. 
Especially, we now show that the
last one yields values consistent with the observed $J's$.

Let us determine the (hopefully realistic)
values of $U$, $U'$, $J_H$, $J_H'$ and $\epsilon$ providing a
$D$ ratio in the needed range (\ref{inequal}). 
To this purpose we consider typical values of $U=10$ eV,
$U'=U-2J_H=6-10$ eV, and $\epsilon =0.1-0.5$ eV,  and we obtain
the ratio $D$ as a function of $J_H/U$. For simplicity the ratio 
$J_H/ J_H'$ is taken equal to 1. Then,
from Fig. 1, we consider  typical ranges
for $D$ between, e.g., $\alpha'^2\approx 1.4$
and $\alpha^2 \approx 4.$, roughly corresponding to
$Q_2/ |Q_1|\approx 6.69$, and $\alpha'^2\approx 2.$
and $\alpha^2 \approx 2.4$, roughly corresponding to
$Q_2/ |Q_1|\approx 2.84$. Then we look for what range
of $J_H/U$ the conditions $4>D>1.4$ or $2.4>D>2.$
are realized. In the first case the quite reasonable 
range $0.1\lesssim J_H/U \lesssim 0.35$ is found,
while in the second case $0.1 \lesssim J_H/U \lesssim 0.15$. 
 The observed values \cite{moussa} $J=0.83$ meV (ferro) and $J'=0.58$ meV
(antiferro) are obtained with $J_H=J'_H=1.25$ eV and $t=0.34$eV in the case
of $Q_2/|Q_1|=2.84$. This value of $t$ is fairly large, but it can
be substantially decreased by taking larger values of  $Q_2/|Q_1|$.
Notice that, contrarily to
the calculations of KK, the planar ferromagnetic coupling can easily be larger 
than the $z$-axis antiferromagnetic one. 
This comes from the ratio $t_{aa}^2/{t'_{aa}}^2$ in 
(\ref{JJ}) where, due to
the orbital order stabilized by the JT distortion,
hopping in the plane directions is enhanced with respect to the $z$-axis.

In conclusion, we have illustrated the possibility of an alternative 
mechanism for the layered antiferromagnetism of the stoichimetric LMO 
compound. In particular, we showed that the superexchange mechanism,
together with strong JT planar distortions, can be responsible for
the specific A-type magnetic structure. Like in the analysis
of Ref. \cite{KK}, the sinergetic effect of both magnetic superexchange
and  orbital ordering is a crucial ingredient.
However, contrary to the assumption of Ref.\cite{KK},
in the present scenario, 
we assumed a given orbital ordering strongly lifting the
degeneracy of the $e_g$ orbitals ($\epsilon \gg J_{ab},J_{aa}$).
A relevant role here is also played by the $t_{2g}$ spin degrees of
freedom, as seen from the expressions (\ref{jaa}) and (\ref{jab}), which
depend rather strongly on $J_H'$.

Despite the basic differences between our scheme and the proposal
of Ref. \cite{KK}, we find some similarities in the overall result.
Specifically we find that for systems with positive $Q_1$,
like, e.g., ${\rm KCuF_3}$, no restriction is needed on the
$Q_2/Q_1$ ratio to make the orbital ordering compatible with 
A-AFMI magnetic structure. This is not the case for negative
$Q_1$, where the above discussed conditions have to be
imposed on $Q_1/|Q_2|$ and, consequently on the values of $J_H/U$. 
Similar values of the $J_H/U$ were found in \cite{KK}.

Within the present model, to be more realistic 
one should also take into account the 
existence of tilt distortions and 
consider the effects of non vanishing transfer
integrals $t'_{xx}$ and $t'_{xz}$. In fact these hopping constants are 
strictly zero only for lattices without tilting of the ${\rm MnO_6}$ octaedra
around axes on the $xy$ plane. We have found that positive $t'_{xz}$ and 
negative $t'_{xx}$ favour $A$-type ordering.
Another contribution to the problem is the antiferromagnetic exchange 
originating from $t_{2g}$ electrons.
Finally, a full calculation involving all superexchange processes should be 
feasible.

We stress again that the mechanism proposed here is based on
the tight interplay of lattice and 
electronic degrees of freedom existing even in the undoped LMO system, and 
that it is aimed to correlate
the magnetic ordering with the JT distortions. Justifying the values of $Q_1$ 
and $Q_2$ from microscopic
grounds is beyond the scope of the present paper. 
Some mixing between lattice and spin dynamics and isotopic
or pressure dependence of the spin-wave velocity are 
expected to be rather natural consequences of the proposed scenario.

%  {\bf%%%%%%%%%%%%%%%%%%%%%%%%%%%%%%%%% ACKNOWLEDGMENTS
%
\acknowledgments 
M.G. acknowledges financial support
of the Istituto Nazionale di Fisica della Materia, Progetto
di Ricerca Avanzata 1996, as well as from 
Universit\'e Joseph Fourier and LEPES,
CNRS in Grenoble where part of this work was carried out.
D.F. acknowledges hospitality and support from the Dipartimento
di Fisica of Universit\`a di Roma ``La Sapienza''.

%
%%%%%%%%%%%%%%%%%%%%%%%%%%%%%%%%% REFERENCES 
%

\end{multicols}

\begin{references}
\bibitem{notaU} 
This is consistent 
with the observation that, in the absence of strong-correlation effects,
the ferromagnetically ordered ${\rm Mn O_2}$
planes at quarter filling (one electron and two $e_g$ orbitals per
site) should be metallic (unless an exceedingly
large JT splitting $\epsilon$ of the order of the bandwidth or more 
is assumed).
\end{references}
\end{document}